\documentclass[aps,superscriptaddress,showpacs,nofootinbib,eqsecnum]{revtex4}

\usepackage{amssymb}
\usepackage{epsf,epsfig}

\begin{document}

\title{Local Approximations for Effective Scalar Field Equations of Motion}

\author{Arjun Berera} \email{ab@ph.ed.ac.uk} \affiliation{School of Physics,
  University of Edinburgh, Edinburgh, EH9 3JZ, United Kingdom}

\author{Ian G. Moss} \email{ian.moss@ncl.ac.uk} \affiliation{School of
  Mathematics and Statistics, University of Newcastle Upon Tyne, NE1 7RU,
  United Kingdom}

\author{Rudnei O. Ramos} \email{rudnei@uerj.br} \affiliation{Departamento de
  F\'{\i}sica Te\'orica, Universidade do Estado do Rio de Janeiro, 20550-013
  Rio de Janeiro, RJ, Brazil}

\begin{abstract}
  
  {}Fluctuation and dissipation dynamics is examined at all temperature ranges
  for the general case of a background time evolving scalar field coupled to
  heavy intermediate quantum fields which in turn are coupled to light quantum
  fields. The evolution of the background field induces particle production
  from the light fields through the action of the intermediate catalyzing
  heavy fields.  Such field configurations are generically present in most
  particle physics models, including Grand Unified and Supersymmetry theories,
  with application of this mechanism possible in inflation, heavy ion
  collision and phase transition dynamics.  The effective evolution equation
  for the background field is obtained and a fluctuation-dissipation theorem
  is derived for this system.  The effective evolution in general is nonlocal
  in time.  Appropriate conditions are found for when these time nonlocal
  effects can be approximated by local terms.  Here careful distinction is
  made between a local expansion and the special case of a derivative
  expansion to all orders, which requires analytic behavior of the evolution
  equation in {}Fourier space.

\end{abstract}


\maketitle

\section{Introduction}

Dissipation physics can play an important role in systems governed by quantum
fields in subject areas as diverse as condensed matter to particle physics and
early universe cosmology.  Dissipation effects can have experimental
consequences in many systems. In cosmology, inflation could occur concurrently
with particle production due to dissipation from the scalar inflaton field in
the context of the warm inflation scenario \cite{im,wi}, leading to observable
signatures in density perturbations \cite{tb,hmb}.  In heavy ion collisions,
the chiral symmetry QCD transition could be accompanied by dissipative effects
\cite{xg}.  More generally, any phase transition in a condensed matter,
particle physics or cosmological system will be accompanied by dissipative
effects during the dynamical evolution of the order parameter characterizing
the phase transition \cite{bv}.

The first principles origin of dissipation from quantum field theory has been
a subject of interest for several decades.  The typical system studied
\cite{hs,morikawa,ring1,boya,GR,bgr,bgr2} is where a background scalar field
is coupled to some variety of bosonic and fermionic light quantum fields, with
the entire system starting off in some statistical state.  The general outcome
is the time evolution of the scalar field induces production of particles from
the light quantum fields, creating dissipative and fluctuating effects.  Such
dynamics is of interest in particle physics, since this direct dissipative
mechanism arises from very generic coupling schemes and also because such
dynamical situations commonly present themselves in problems such as
inflationary cosmology, heavy ion physics and in phase transitions in general.

In particle physics models there is a novel interaction configuration which
also has been shown to lead to considerable dissipative effects, and that is
the case where a background scalar field is coupled to heavy intermediate
quantum fields which in turn are coupled to the light quantum fields
\cite{br,br2}.  In this two-stage mechanism, the background scalar field
indirectly induces particle production in the light fields through the
intermediate heavy fields which in a sense help to catalyze the effect.
Dissipation from such field configurations in systems near thermal equilibrium
has been shown to arise \cite{mx} and application of these effects to
inflation have been shown to have significant importance \cite{bb4}.

A full understanding of this mechanism requires understanding not only the
dissipative effects but also the fluctuations created and the relation between
the two in the context of a fluctuation-dissipation theorem.  This so far has
not been done for this two-stage mechanism and will be one of the results of
this paper.

To arrive at this result, what will be required is to obtain the effective
evolution equation for the background scalar field once all other fields to
which it is coupled are integrated out.  The resulting effective evolution
equation, which will be derived using the Schwinger closed-time formalism
\cite{schw}, in general will have time nonlocal terms.  The second interest of
this paper is in determining when such nonlocal terms can be well approximated
by local terms. In general this requires a careful examination of the
different time scales involved in the dynamics of the system in study to
determine the importance of nonlocal effects.  {}For instance, when the
characteristic time scale of the nonlocal (or memory) terms in the effective
equation of motion is close to the typical reaction time of the system, it has
been shown that large effects on the dynamics can arise as a consequence of
the nonlocal terms as compared to the results coming from a local
approximation for the equations \cite{KKR}. In this paper we adopted the
general approach of directly analyzing the nonlocal terms and then determining
the relevant time scale for when we can approximate them as local forms.

In practical application of these types of equations, there is an immense
saving of effort as well as much more transparent understanding of the physics
from a local equation as opposed to a nonlocal one, since the former can
generally be analyzed with much less numerical treatment than the latter.
Thus it is a very important question when and how accurately the generally
nonlocal effective equations can be approximated by a local form.  One very
special type of local approximation is the derivative expansion
\cite{hs,GR,dh,bgr,bgr2,im2}.  This is just a Taylor expansion of the nonlocal
terms about some time point.  Such an expansion therefore requires analytic
behavior of the relevant nonlocal terms.  However that may not arise in all
circumstances, and for example even simple logarithmic corrections would make
the derivative expansion invalid.  Nevertheless, a careful distinction must be
made between localized behavior as opposed to localized analytic behavior.
Even in the general circumstance where the behavior is not analytic, the
nonlocal terms may still be fairly peaked functions that can be well
approximated as local.  The nonlocal terms in the effective evolution equation
we derive for this two-stage mechanism will be examined for when they can be
approximated as localized.  Although in this paper we carry out this
investigation in the context of the two-stage mechanism, most of our
significant results here would also apply to the direct dissipative mechanism
(for a thorough discussion about the two-stage decay mechanism contrasted to
the direct one, see e.g. Ref. \cite{br,br2}).

We assume throughout that the system is enclosed by an environment which is
initially close to thermal equilibrium and remains in that state due to the
existence of a relatively short interaction timescale in the various
components contributing to the environment. In this respect we have a
different situation than the one considered in the original work on the two
stage mechanism \cite{br,br2} and more recent numerical work \cite{aarts}. In
fact, we find that the {\it local} dissipative effects cease in the limit that
the temperature $T\to 0$, as one might expect and in agreement with
\cite{mx,aarts}.

This paper is organized as follows. In Sec. II we give the Lagrangian density
model used to study the two-stage decay mechanism and the respective effective
equation of motion for a background scalar field that exhibits both the
characteristic dissipation and fluctuation (stochastic) like terms. In Sec.
III we derive the various generalized fluctuation-dissipation relations for
the model under study and show the generality of these relations. In Sec. IV
we define the local limit for the noise and dissipation terms and then we
provide a thorough study both analytically and numerical to support
localization in time for the noise and dissipation kernels appearing in the
effective equation of motion given in Sec. II. {}Finally, our concluding
remarks are given in Sec. V.

\section{The Model}

We consider the following model of a scalar field $\Phi$ interacting
indirectly through a catalyst field $\chi$ to a scalar field $\sigma$, which
we can think of as our heat bath. The field $\Phi$ has no direct coupling
terms to the field $\sigma$. The Lagrangian density for this model is the
following

\begin{eqnarray} 
{\cal L} [ \Phi, \chi, \sigma] &=&  
\frac{1}{2} \partial_\mu \Phi \partial_\nu \Phi - \frac{m_\phi^2}{2}\Phi^2 - 
\frac{\lambda}{4 !} \Phi^4  
\nonumber \\
&+&   \frac{1}{2} \partial_\mu \chi \partial_\nu \chi - 
\frac{m_{\chi}^2}{2}\chi^2 
- \frac{f}{4!} \chi^4 - \frac{g^2}{2} 
\Phi^2 \chi^2  
\nonumber \\ 
&+& \frac{1}{2} \partial_\mu \sigma \partial_\nu \sigma - 
\frac{m_\sigma^2}{2}\sigma^2 - 
h {\cal M} \chi \sigma^2 
\: ,
\label{model} 
\end{eqnarray} 
where ${\cal M}$ is a parameter with mass dimension added in order to make the
coupling constant $h$ dimensionless.

We consider $\Phi(x) = \varphi_c(x) + \phi(x)$ where $\varphi_c(x)$ is a
background like component of the scalar field, or vacuum expectation value of
$\Phi$, and $\phi(x)$ are the quantum fluctuations around this background
field. The nonequilibrium equation of motion for $\varphi_c$ can be derived
using for example the functional closed time path (CTP) formulation of
Schwinger-Keldysh \cite{schw}. In the CTP formalism the time integration is
along a contour $c$ from $- \infty$ to $+ \infty$ and then back to $-\infty$.
{}For reviews please see, for example, Refs. \cite{chou,rivers,weert}. The
detailed derivation of the effective equation of motion (EOM) for the field
$\varphi_c$ was explained in many previous papers (see e.g.
\cite{GR,br,br2,yoko}) and therefore we will only give here the final result
for it. The final result is expressed in terms of effective propagators (that
already include the effects of the bath fields).  In the CTP formalism these
field propagators, e.g. for $\phi$, are given by \cite{GR} (which also apply
to any of the other boson fields):

\begin{eqnarray}
&&G_\phi^{++}(x,x') = i \langle T_{+} \phi(x) \phi(x')\rangle\:,
\nonumber \\
& & G_\phi^{--}(x,x') = i \langle T_{-} \phi(x) \phi(x')\rangle\:,
\nonumber \\
& & G_\phi^{+-}(x,x') = i \langle \phi(x') \phi(x)\rangle\:,
\nonumber \\
& & G_\phi^{-+}(x,x') = i \langle \phi(x) \phi(x')\rangle \: ,
\label{twopoint}
\end{eqnarray}

\noindent
where $T_{+}$ and $T_{-}$ indicate chronological and anti-chronological
ordering, respectively.  $G_\phi^{++}$ is the usual Feynman propagator. The
other three propagators come as a consequence of the time contour and are
considered as auxiliary (unphysical) propagators. {}From (\ref{twopoint}) it
also follows that the boson field propagators can also be expressed as

\begin{eqnarray}
&&G^{++}_\phi(x,x') = G^{>}_\phi(x,x')
\theta(t-t') + G^{<}_\phi(x,x') \theta(t'-t) ,
\nonumber \\
& & G^{--}_\phi(x,x') = G^{>}_\phi(x,x')
\theta(t'-t) + G^{<}_\phi(x,x') \theta(t-t') ,
\nonumber \\
& & G^{+-}_\phi(x,x') = G^{<}_\phi(x,x') ,
\nonumber \\
& & G^{-+}_\phi(x,x') = G^{>}_\phi(x,x')\;.
\label{Gs}
\end{eqnarray}

In deriving the effective EOM for $\varphi_c$ we first integrate over the
$\sigma$ field. In this first step the Green's functions for the fields (in
this case for the $\chi$ field in leading order) coupled to the fermions get
dressed by $\sigma$ field corrections. Subsequently functional integration
over the other fields (in our case the $\chi$ and $\phi$ field fluctuations)
is performed.  In this process, we can express the boson Green's functions
entering in the $\varphi_c$ effective EOM in the generic form

\begin{eqnarray}
\left[\frac{\partial^2}{\partial t^2} 
-\nabla^2 + M^2 \right] 
G (x,x')
+ \int d^4 z \, \Sigma (x,z) \, G (z,x') = 
\delta (x,x') \;,
\label{G}
\end{eqnarray}
where $G(x,x')$ and $\Sigma(x,x')$ are the (matrix) propagator and
self-energy, respectively. The mass term $M^2$ in the above equation is the
one appropriate the the respective field.

In terms of a loop expansion for the fluctuation field $\phi$ and $\chi$ and
the generic expressions for the field propagators given above, the effective
equation of motion that emerges for $\varphi_c$ is of the form
\cite{GR,br,br2,yoko}

\begin{eqnarray}
\lefteqn{\left[ \Box + M_\phi^2 \right]\varphi_c (x) +
\frac{\lambda}{3 !} \varphi_c^3 (x) }\nonumber \\
& & 
+\varphi_c (x) \int d^4 x'
\varphi_c^2 (x') \left\{ \frac{\lambda^2}{2} 
{\rm Im}\left[G_\phi^{++}(x,x')\right]^2
+  2 g^4 
{\rm Im} \left[G_{\chi_j}^{++}(x,x')\right]^2 \right\} \theta(t-t')
\nonumber \\
&&+ \int d^4 x'\varphi_c (x')  \left\{ \frac{\lambda^2}{3}  
{\rm Im}\left[G_\phi^{++}(x,x')\right]^3 + 4 g^4 
{\rm Im}\left[G_\chi^{++}(x,x')G_\phi^{++}(x,x')G_\chi^{++}(x,x')\right]
 \right\} \theta(t-t') \nonumber \\
& &  = \varphi_c (x) \xi_1 (x) + \xi_2 (x) \: ,
\label{full eqmotion}
\end{eqnarray}
where $M_\phi$ denotes the renormalized effective mass for $\phi$ that
includes all local contributions. The nonlocal contributions can be associated
to dissipation and noise terms. {}For instance, the  third and forth terms in Eq. 
(\ref{full eqmotion}) are related to dissipation terms. The
source terms $\xi_1$ and $\xi_2$ are stochastic (noise) fields associated with
the imaginary terms in the effective action coming from the real-time
evaluation of the perturbative loop diagrams leading to Eq (\ref{full
  eqmotion}).  They have two-point correlation functions given by \cite{GR}

\begin{equation}
\langle \xi_1 (x) \xi_1 (x')\rangle = \frac{\lambda^2}{2}
{\rm Re}\left[G_\phi^{++}(x,x')\right]^2  + 
2 g^4 {\rm Re}\left[G_{\chi}^{++}(x,x')\right]^2\: ,
\label{noise1}
\end{equation}
and

\begin{equation}
\langle \xi_2 (x) \xi_2 (x')\rangle = \frac{\lambda^2}{6}
{\rm Re}\left[G_\phi^{++}(x,x')\right]^3  + 
2 g^4 {\rm Re}\left[G_\chi^{++}(x,x')^2G_{\phi}^{++}(x,x')\right]\: ,
\label{noise2}
\end{equation}
and are colored (space-time dependent) in general and Gaussian distributed,
with probability distributions given by ($N_1$ and $N_2$ are appropriate
normalization constants)

\begin{equation}
P[\xi_1] = N_1^{-1} \exp\left\{ - \frac{1}{2} \int d^4 x d^4 x'
\xi_1 (x) \left[ \frac{\lambda^2}{2} {\rm Re}
\left[G_\phi^{++}(x,x')\right]^2
+ 2 \sum _j g_j^4 {\rm Re} \left[G_{\chi_j}^{++}(x,x')\right]^2
\right]^{-1} \xi_1 (x')
\right\} \:,
\label{P1}
\end{equation}
and

\begin{equation}
P[\xi_2] = N_2^{-1} \exp\left\{ - \frac{1}{2} \int d^4 x d^4 x'
\xi_2 (x) \left[ \frac{\lambda^2}{6}
{\rm Re}\left[G_\phi^{++}(x,x')\right]^3  + 
2 g^4 {\rm Re}\left[G_\chi^{++}(x,x')^2G_{\phi}^{++}(x,x')\right]
\right]^{-1} \xi_2 (x')
\right\} \:.
\label{P2}
\end{equation}
Both dissipation and noise kernel are related to generalized
fluctuation-dissipation relations, as we show next.

\section{Fluctuation-Dissipation Relations}

A generalized fluctuation-dissipation relation between the nonlocal kernels in
Eq. (\ref{full eqmotion}) and the stochastic fields correlation functions were
derived in Ref. \cite{GR}, while explicit expressions derived from the
dissipation and noise kernels appearing in (\ref{full eqmotion}) were obtained
by Yokoyama in Ref.  \cite{yoko}.  General expressions that relate both
kernels can also easily be derived in the following way. Lets first define
dissipative kernels, ${\cal D}_i$, as given by

\begin{eqnarray}
{\cal C}_i({\bf x}-{\bf x'},t-t') = - \frac{\partial}{\partial t'}
{\cal D}_i({\bf x}-{\bf x'},t-t')\:,
\label{dissip kernels}
\end{eqnarray}
where the ${\cal C}_{i=1,2}$ are defined by

\begin{eqnarray}
{\cal C}_{1}({\bf x}-{\bf x'},t-t') = \lambda^2
{\rm Im}\left[G_\phi^{++}(x,x')\right]^2 {\rm sgn} (t-t')
+  4 g^4 
{\rm Im} \left[G_{\chi_j}^{++}(x,x')\right]^2 {\rm sgn}(t-t') \:,
\label{C1}
\end{eqnarray}
and

\begin{eqnarray}
{\cal C}_{2}({\bf x}-{\bf x'},t-t') = \frac{\lambda^2}{3}  
{\rm Im}\left[G_\phi^{++}(x,x')\right]^3 {\rm sgn}(t-t') + 4 g^4 
{\rm Im}\left[G_\chi^{++}(x,x')G_\phi^{++}(x,x')G_\chi^{++}(x,x')\right]
{\rm sgn}(t-t')\:,
\label{C2}
\end{eqnarray}

\noindent
where ${\rm sgn}(t-t') = \theta(t-t') - \theta(t'-t)$ is the sign function
(note that only the part for $t>t'$ of these functions enters in the equation
of motion Eq. (\ref{full eqmotion})).  The extra factor 2 in the definition of
Eq. (\ref{C1}) is for convention purposes, such that both ${\cal C}_{i=1,2}$
will satisfy the same fluctuation-dissipation relation.  Using Eqs. (\ref{C1})
and (\ref{C2}) in (\ref{full eqmotion}) it follows that

\begin{eqnarray}
\lefteqn{\left[ \Box + M_\phi^2\right]\varphi_c (x) +
\frac{\lambda}{3 !} \varphi_c^3 (x) }\nonumber \\
& & 
-\frac{\varphi_c (x)}{2} \int d^3 x'
\varphi_c^2 ({\bf x}',t) {\cal D}_1({\bf x}-{\bf x'},0) 
- \int d^3 x'
\varphi_c ({\bf x}',t) {\cal D}_2({\bf x}-{\bf x'},0) 
\nonumber \\
&&+  \varphi_c (x) \int d^3 x' \int_{-\infty}^t dt'
\varphi_c ({\bf x'},t') \dot{\varphi_c} ({\bf x'},t') {\cal D}_1({\bf x}-{\bf
x'},t-t')
+  \int d^3 x' \int_{-\infty}^t dt'
\dot{\varphi_c} ({\bf x'},t') {\cal D}_2({\bf x}-{\bf x'},t-t')
\nonumber \\
& & = \varphi_c (x) \xi_1 (x) + \xi_2 (x) \: .
\label{full eqmotion2}
\end{eqnarray}

\noindent
The third and forth terms in Eq. (\ref{full eqmotion2}) can be
shown to correspond, in the homogeneous approximation for the field
$\varphi_c$, to the one-loop $\Phi$ vertex correction and the two-loop setting
sun like mass corrections, respectively, for the model (\ref{model}). Its is
clear now from (\ref{full eqmotion2}) that ${\cal D}_i$ are dissipative like
terms. That they satisfy generalized fluctuation-dissipation relations with
the noise kernels defined by Eqs. (\ref{noise1}) and (\ref{noise2}) can be
demonstrated when space-time Fourier transforming both kernels.  We can write
the {}Fourier transformed kernel ${\cal F}[{\cal C}_i({\bf x}-{\bf x}',t-t')]
= \tilde{{\cal C}}_i ({\bf p}, \omega)$, where the Fourier transform ${\cal
  F}[ \cdots ]$ is defined by

\begin{equation}
{\cal F}[ \cdots ] = \int d^3 x \, dt  \: [\cdots]\:
e^{ -i {\bf p} \cdot {\bf x} + i \omega t} \;.
\end{equation}
Note that, from the Eqs. (\ref{C1}) and (\ref{C2}), the kernels ${\cal C}_i$
are anti-symmetrical in the time.  It then follows that their time {}Fourier
transforms are purely imaginary. Thus, we can write \cite{yoko}

\begin{equation}
\tilde{{\cal C}}_i ({\bf p}, \omega) = 
- 2 i \omega \tilde{\Gamma}_i ({\bf p}, \omega)\;,
\label{Comega}
\end{equation}
where $\tilde{\Gamma}_i ({\bf p}, \omega)$ is a real quantity.  The
transformed dissipation kernels become

\begin{equation}
\tilde{{\cal D}}_i ({\bf p}, \omega) =  
2 \tilde{\Gamma}_i ({\bf p}, \omega)\;.
\label{Domega}
\end{equation}
Similarly, for the noise terms we introduce the kernels ${\cal N}_i({\bf
  x}-{\bf x'},t-t')$, where from Eqs. (\ref{noise1}) and (\ref{noise2}), they
are defined by

\begin{eqnarray}
{\cal N}_{1}({\bf x}-{\bf x'},t-t') = \frac{\lambda^2}{2}
{\rm Re}\left[G_\phi^{++}(x,x')\right]^2  + 
2 g^4 {\rm Re}\left[G_{\chi}^{++}(x,x')\right]^2 \:,
\label{N1}
\end{eqnarray}
and

\begin{eqnarray}
{\cal N}_{2}({\bf x}-{\bf x'},t-t') =    \frac{\lambda^2}{6}
{\rm Re}\left[G_\phi^{++}(x,x')\right]^3  + 
2 g^4 {\rm Re}\left[G_\chi^{++}(x,x')^2G_{\phi}^{++}(x,x')\right]\:,
\label{N2}
\end{eqnarray}

\noindent
and using again the properties of the field propagators in the CTP formalism,
we notice that ${\cal N}_i$ are symmetrical in both space and time.  Thus,
their {}Fourier transforms in space and time, ${\cal F}[ {\cal N}_i (x,x') ] =
\tilde{{\cal N}}_i ({\bf p}, \omega)$, are purely real.

{}From this point on, we will restrict ourselves to parameter values such that
$\lambda = {\cal O}(g^4)$ and $g^2 \ll 1$. In this situation we can drop the
first term in the right-hand-side of both Eqs. (\ref{N1}) and (\ref{N2}) and
likewise in the expressions for the kernels (\ref{C1}) and (\ref{C2}).
Repeating the analysis with these terms included presents no specific
difficulties, but it is not particularly illuminating.

In order to relate the dissipation and noise kernels, it is convenient to
write the Feynman propagator $G^{++}(x,x')$ in real and imaginary parts as
\cite{gert}
\begin{equation}
G^{++}(x,x') = F(x,x') + \frac{1}{2} \rho(x,x') {\rm sgn}(t-t')\;,
\label{GFrho}
\end{equation}
where $F(x,x')$ and $\rho(x,x')$ are the anti-commutator and commutator (or
the statistical and spectral functions) of two fields defined, respectively,
by

\begin{equation}
{}F(x,x') =  \frac{i}{2} \langle \{ \phi(x),\phi(x')\} \rangle
= \frac{1}{2} \left[G^<(x,x') + G^>(x,x') \right]\;,
\label{F}
\end{equation}
and

\begin{equation}
\rho(x,x')  =  i \langle [\phi(x),\phi(x') ] \rangle
=  \left[G^>(x,x') - G^<(x,x') \right]\;.
\label{rho0}
\end{equation}
{}For real fields, $F(x,x')$ and $\rho(x,x')$ are real and pure imaginary
functions respectively, and have the symmetry properties $F(x,x') = F(x',x)$
and $\rho(x,x')=-\rho(x',x)$, which easily follow from the definitions
(\ref{twopoint}) and (\ref{Gs}). Note also that since $\rho(x,x')$ is an
anti-symmetrical function, its Fourier transform, ${\cal F}[ \rho(x,x')
]=\tilde\rho$, is real.

In terms of (\ref{F}) and (\ref{rho0}), the noise kernel ${\cal N}_1$ becomes

\begin{eqnarray}
{\cal N}_1({\bf x}-{\bf x}',t-t') &=& 2 g^4 
{\rm Re}\left[G_\chi^{++}(x,x')\right]^2
\nonumber \\
&=& 2 g^4\left[ F_\chi({\bf x}-{\bf x}',t-t')^2  + 
{1\over 4} \rho_\chi({\bf x}-{\bf x}',t-t')^2 \right] \;,
\end{eqnarray}
or in terms of {}Fourier transforms,

\begin{eqnarray}
{\cal F}[ {\cal N}_1({\bf x}-{\bf x}',t-t') ] 
&=& \tilde{{\cal N}}_1({\bf p}, \omega) \nonumber \\
&=& 
2g^4 \int \frac{d^3 k}{(2 \pi)^3} \int_{-\infty}^\infty {d\omega' \over 2\pi}
\left[ \tilde{F}_\chi({\bf k},\omega') 
\tilde{F}_\chi({\bf p}-{\bf k},\omega-\omega')+
{1\over 4} \tilde{\rho}_\chi ({\bf k},\omega') 
\tilde{\rho}_\chi({\bf p}-{\bf k},\omega-\omega')\right].
\label{N1pomega}
\end{eqnarray}

If we now use the property valid at equilibrium, coming from the
Kubo-Martin-Schwinger relation for the propagators, $G^>({\bf p}, \omega) =
\exp(\beta \omega) G^<({\bf p}, \omega)$, it follows, from the definition of
the spectral density, that \cite{lebellac}

\begin{eqnarray}
&&G^{>}({\bf p}, \omega) = 
\left[ 1 + n(\omega) \right] \tilde{\rho}({\bf p}, \omega) \;,
\nonumber \\
&&G^{<}({\bf p}, \omega) = n(\omega) \tilde{\rho}({\bf p}, \omega) \;,
\label{Grho}
\end{eqnarray}
where $n(\omega)$ is the Bose-Einstein distribution function,

\begin{equation}
n(\omega) = \frac{1}{e^{\beta \omega} -1}\:,
\end{equation}
and then we obtain for the {}Fourier transform of the statistical function,
${\cal F}[ {}F(x,x') ]= \tilde{F}({\bf p}, \omega)$, the result

\begin{equation}
\tilde{F}({\bf p}, \omega) =\frac{1}{2} [1+2n(\omega)] 
\tilde{\rho}({\bf p}, \omega)\;.
\label{Frho}
\end{equation}

Using Eq. (\ref{Frho}) and the identity
\begin{equation}
1+[1+2n(\omega')][1+2n(\omega-\omega')]=2\left( e^{\beta\omega}+1 \right)
n(\omega')n(\omega-\omega')  \;,
\label{identity}
\end{equation} 
we obtain for Eq. (\ref{N1pomega}) the result

\begin{equation}
\tilde{{\cal N}}_1({\bf p},\omega)=g^4\left(e^{\beta\omega}+1\right)
\int \frac{d^3 k}{(2 \pi)^3} \int_{-\infty}^\infty {d\omega' \over 2\pi}
n(\omega') n(\omega-\omega')
\tilde{\rho}_\chi({\bf k},\omega') 
\tilde{\rho}_\chi({\bf p}-{\bf k},\omega-\omega')\;.
\label{noise1kernel}
\end{equation}
Analogously, for the kernel ${\cal C}_1({\bf x}-{\bf x}',t-t')$, we find

\begin{eqnarray}
{\cal C}_1({\bf x}-{\bf x}',t-t') &=& 
4 g^4{\rm Im}\left[G_\chi^{++}(x,x')\right]^2
{\rm sgn}(t-t')
\nonumber \\
&=& 4i g^4 F_\chi({\bf x}-{\bf x}',t-t') \rho_\chi({\bf x}-{\bf x}',t-t') \;,
\end{eqnarray}
and expressing it in terms of transforms, after some algebra, we find

\begin{eqnarray}
\tilde{{\cal C}}_1({\bf p}, \omega) = -2 i g^4 
\left(e^{\beta\omega} - 1 \right)
\int \frac{d^3 k}{(2 \pi)^3} \int_{-\infty}^\infty {d\omega' \over 2\pi}
n(\omega') n(\omega-\omega')
\tilde{\rho}_\chi({\bf k},\omega') 
\tilde{\rho}_\chi({\bf p}-{\bf k},\omega-\omega')\;.
\label{C1kernel}
\end{eqnarray}
Using now Eq. (\ref{Comega}) we finally find that the dissipation and noise
kernel satisfy the generalized fluctuation-dissipation relation

\begin{equation}
\tilde{{\cal N}}_1 ({\bf p}, \omega) = 2 \omega \left[ n(\omega) +
\frac{1}{2} \right] 
\tilde{\Gamma}_1 ({\bf p}, \omega)  \;.
\label{fluct-diss}
\end{equation}

Likewise, for the noise kernel ${\cal N}_2$ we find
\begin{eqnarray}
{\cal N}_2({\bf x}-{\bf x}',t-t') &=&2 g^4 {\rm Re}\left[G_\chi^{++}(x,x')^2
G_\phi^{++}(x,x')\right]
\nonumber \\
&=& 2 g^4 \left\{
\left[ F_\chi({\bf x}-{\bf x}',t-t')^2  
+{1\over 4}
\rho_\chi({\bf x}-{\bf x}',t-t')^2 \right]  F_\phi({\bf x}-{\bf x}',t-t') 
\right. \nonumber \\
& + & \left. \frac{1}{2} F_\chi({\bf x}-{\bf x}',t-t')
\rho_\chi({\bf x}-{\bf x}',t-t') \rho_\phi({\bf x}-{\bf x}',t-t') 
\right\}\;,
\end{eqnarray}
and in terms of {}Fourier transforms,

\begin{eqnarray}
\tilde{{\cal N}}_2({\bf p},\omega)
&=& g^4\left(e^{\beta\omega}+1\right)
\int \frac{d^3 k}{(2 \pi)^3} \frac{d^3 q}{(2 \pi)^3}
\int_{-\infty}^\infty {d\omega_1 \over 2\pi}{d\omega_2 \over 2\pi}
n(\omega_1)n(\omega_2) n(\omega-\omega_1-\omega_2) \nonumber \\
&\times& 
\tilde{\rho}_\chi({\bf k},\omega_1) \tilde{\rho}_\chi({\bf q},\omega_2)
\tilde{\rho}_\phi({\bf p}-{\bf k}-{\bf q},\omega-\omega_1-\omega_2)\;,
\label{noise2kernel}
\end{eqnarray}
while for the transform of the kernel ${\cal C}_2({\bf x}-{\bf x}',t-t')$ we
find

\begin{eqnarray}
\tilde{{\cal C}}_2({\bf p},\omega)
&=& - 2 i g^4\left(e^{\beta\omega}-1\right)
\int \frac{d^3 k}{(2 \pi)^3} \frac{d^3 q}{(2 \pi)^3}
\int_{-\infty}^\infty {d\omega_1 \over 2\pi}{d\omega_2 \over 2\pi}
n(\omega_1)n(\omega_2) n(\omega-\omega_1-\omega_2) \nonumber \\
&\times& 
\tilde{\rho}_\chi({\bf k},\omega_1) \tilde{\rho}_\chi({\bf q},\omega_2)
\tilde{\rho}_\phi({\bf p}-{\bf k}-{\bf q},\omega-\omega_1-\omega_2)\;.
\label{C2kernel}
\end{eqnarray}
Upon using Eq. (\ref{Comega}), we again find that the dissipation and noise
kernels $\Gamma_2({\bf p},\omega)$ and $\tilde{{\cal N}}_2({\bf p},\omega)$
are also related in an identical way to Eq. (\ref{fluct-diss}).

These fluctuation-dissipation relations are actually generic to any product of
two-point Green functions in the CTP formalism.  To see this, begin by
expressing the self-energy matrix elements in a similar way to the CTP
propagators as

\begin{eqnarray}
&& \Sigma^{++}(x,x') = \Sigma^{>}(x,x')
\theta(t-t') + \Sigma^{<}(x,x') \theta(t'-t) ,
\nonumber \\
& & \Sigma^{--}(x,x') = \Sigma^{>}(x,x')
\theta(t'-t) + \Sigma^{<}(x,x') \theta(t-t') ,
\nonumber \\
& & \Sigma^{+-}(x,x') =- \Sigma^{<}(x,x') ,
\nonumber \\
& & \Sigma^{-+}(x,x') = -\Sigma^{>}(x,x')\; ,
\label{Sigma}
\end{eqnarray}
{}From (\ref{Sigma}) the following property follows,
\begin{equation}
\Sigma^{++}(x,x') + \Sigma^{+-}(x,x') + \Sigma^{-+}(x,x') +
\Sigma^{--}(x,x') =0\;.
\label{sum Sigma}
\end{equation}
In addition, the self-energies in (\ref{Sigma}) are found in general to
satisfy conditions similar to those for the propagators, such as

\begin{eqnarray}
&&\Sigma^{>}(x,x') = \Sigma^< (x',x)\;,
\nonumber \\
&&
\left[i\Sigma^{>(<)}(x,x')\right]^\dagger  = i\Sigma^{<(>)} (x,x') \;.
\label{Sigma cond}
\end{eqnarray}

The real and imaginary parts for the self-energy component $\Sigma^{++}(p)$ in
the CTP formalism can be separated by introducing the commutator and
anticommutator functions, analogous to the ones used to express the causal
propagator Eq. (\ref{GFrho}),

\begin{equation}
\Sigma^{++}(x,x')=\Sigma_F(x,x')+{1\over 2}\Sigma_\rho(x,x')
\,{\rm sgn}(t-t')
\;,
\label{Sigma++}
\end{equation}
where

\begin{eqnarray}
&& \Sigma_\rho(x,x')=[\Sigma^{>}(x,x')-\Sigma^{<}(x,x')]\;, \nonumber \\
&& \Sigma_F(x,x')=\frac{1}{2}[\Sigma^{>}(x,x')+\Sigma^{<}(x,x')]\;,
\end{eqnarray}
which are pure imaginary and real respectively. Turning to our earlier
example, Eq. (\ref{full eqmotion}), we find that the dissipation kernels are
related to $\Sigma_\rho$ and the noise kernels to $\Sigma_F$.

{}For the dressed propagators as defined in (\ref{G}), we can express the
transform of the spectral function, $\tilde{\rho}({\bf p},\omega)$, in the
form

\begin{equation}
\tilde{\rho}({\bf p}, \omega) = 
i\left (\omega^2 -\omega({\bf p})^2
+\frac12\Sigma_F +\frac12i\tilde\Sigma_\rho \right)^{-1}
-i\left (\omega^2 -\omega({\bf p})^2
+\frac12\Sigma_F- \frac12i\tilde\Sigma_\rho \right)^{-1} ,
\label{rho}
\end{equation}
where $\omega({\bf p})^2=|{\bf p}|^2+M^2$. The Breit-Wigner form of Eq.
(\ref{rho}) suggests that we can introduce a generalized relaxation time
$\tau({\bf p},\omega)$ of the particle state via the relation

\begin{equation}
\Sigma_\rho({\bf p},\omega) = 4 \omega({\bf p}) \tau({\bf p},\omega)^{-1}.
\end{equation}
The equilibrium self-energies can be expressed in terms of $\Sigma_\rho$,

\begin{eqnarray}
&&\tilde{\Sigma}^{>}({\bf p}, \omega) =  
\left[ 1 + n(\omega) \right] \Sigma_\rho({\bf p},\omega) \;,
\nonumber \\
&&\tilde{\Sigma}^{<}({\bf p},\omega) = 
 n(\omega) \Sigma_\rho ({\bf p},\omega) \;.
\label{Sigma><tilde}
\end{eqnarray}
Using Eq.  (\ref{Sigma><tilde}), we find a local fluctuation dissipation
relationship between the {}Fourier transforms of $\Sigma_F$ and $\Sigma_\rho$,

\begin{equation}
\tilde{\Sigma}_F({\bf p}, \omega)={1\over 2}\left[1+2n(\omega)\right]
\tilde{\Sigma}_\rho({\bf p}, \omega)\;,
\end{equation}
where we have defined ${\cal F}[\Sigma_F(x,x')] = \tilde{\Sigma}_F({\bf p},
\omega)$ and ${\cal F}[\Sigma_\rho(x,x')] = \tilde{\Sigma}_\rho({\bf p},
\omega)$.

\section{Local limit for the noise kernels}

We now study the possible localization of the effective EOM for the background
field.  We first note that nonlinear equations of the form of Eq.  (\ref{full
  eqmotion2}) are impossible to solve analytically.  They are also
notoriously difficult to solve numerically, since they typically exhibit
highly oscillating nonlocal kernels that can lead to errors which
quickly build up and that are too hard to control, thus preventing
any simple numerical attempt to solve these kind of equations.  There are
however different approaches that can be applied to analyze these equations.
{}For instance, in those cases where the kernels exhibit a strong exponential
damping in time, like in the analysis performed in Refs. \cite{br,br2}, a
numerical approach to analyze the nonlocal equation of motion for the background
field is possible and these analysis have shown that a local approximation for
the kernels, and then for the equation of motion, is in very good agreement
with the full numerical solution of the nonlocal equation.  There are however
cases where such a strong damping behavior for the kernels are not possible
(see e.g. Ref. \cite{mx}) and we must resort to other alternative analysis to
determine how good a local approximation is for the nonlocal equation.
For example what we are going to show below is the nonlocal kernel
can be analyzed and used to infer the validity of the local
approximation, in particular determine a time scale for
locality;  all this can be accomplished without reverting to
solving complicated
equations of motion like Eq.  (\ref{full eqmotion2}).
This is our objective here.
Note that since both dissipation and noise kernels are related, if we find
that the noise kernels can be localized, then the same must apply to the
dissipation kernels in the effective EOM. We shall consider homogeneous
background fields, for which the main issue will be the localization in time.
Throughout this section we assume that the relaxation time of the environment
is sufficiently short to be neglected.

{}First of all, we shall define what we mean by a localization of the noise
kernels ${\cal N}$. The concept of localization depends on a particular
context. {}For example, if we are interested in the origin of density
fluctuations in the early universe we might be interested in localization of
the noise kernel on timescales smaller than the Hubble time. We shall
therefore introduce the concept of localization with respect to a particular
timescale $\tau$.

We shall say that a function $f(t)$ is {\em slowly varying on a timescale}
$\tau$ if the Fourier transform $\tilde f(\omega)$ satisfies

\begin{equation}
\tilde f(\omega)=0 \; \hbox{  for  }\; \omega>2\pi/\tau\;.
\end{equation}
A kernel function ${\cal K}(t)$ will be described as {\em localized on a
  timescale $\tau$ with accuracy $1-\epsilon$} if the Fourier transform
$\tilde{\cal K}(\omega)$ satisfies

\begin{equation}
\left|{\tilde{\cal K}(\omega)-\tilde{\cal K}(0)\over \tilde{\cal K}(0)}\right|
<\epsilon \; \hbox{  for  }\; \omega<2\pi/\tau\;.
\label{localdef}
\end{equation}
It follows that, for $f(t)$ slowly varying on a timescale $\tau$ and ${\cal
  K}(t)$ localized on a timescale $\tau$,

\begin{equation}
\int_{-\infty}^\infty {\cal K}(t-t')f(t')dt'=\tilde{\cal K}(0)f(t)+
{\cal O}(\epsilon)\;,
\end{equation}
which is a formal way of stating that ${\cal K}(t)\approx \tilde{\cal
  K}(0)\delta(t-t')$. Note that, if $\tilde{\cal K}(\omega)$ is analytic in
$\omega$, then the kernel can be localized and the kernel admits a local
derivative expansion. In general, the derivative expansion might not exist,
even when the kernel is localized in the sense given above.

To help understand the motivation for the above definition, it may be useful
to think of a Gaussian function $\exp(-t^2/\tau^2_{1/2})$ with width
$\tau_{1/2}$.  The corresponding width of the {}Fourier transform is
$2\pi/\tau_{1/2}$.  By the definition adopted above, on the scale
$\tau_{1/2}$, we would say that the function is not fully localized but
roughly speaking about half localized, thus localized with accuracy around
$50\%$.  {}For the Gaussian to appear localized in t-space, one actually
should look at it from a much bigger scale of say $10\tau_{1/2}$, which means
a much smaller region in $\omega$-space near the maximum part of the gaussian
where it does not vary too much, and thus $\epsilon$ in Eq. (\ref{localdef})
would be small.

As shown in the previous section, the kernels ${\cal C}_j$ are related to the
dissipation kernels ${\cal D}_j$ by Eq. (\ref{dissip kernels}), which
transformed using $\omega$ (we omit the spatial variable for homogeneous
fields) become,

\begin{equation}
\tilde{{\cal D}}_j( \omega) = \frac{i}{\omega} \tilde{{\cal C}}_j( \omega) = 
2 \tilde{\Gamma}_j( \omega)\;.
\label{dissip forms}
\end{equation}
However, from the fluctuation-dissipation relation (\ref{fluct-diss}), we have
that

\begin{equation}
\tilde{{\cal N}}_j(\omega) = 2 \omega \left[ n(\omega) + \frac{1}{2} \right]
\tilde{\Gamma}_j(\omega)\;.
\label{fluctdiss2}
\end{equation}
If a local approximation for the noise two-point correlation function exists,
{\it i.e.} ${\cal N} (t-t') \equiv N_0 \delta(t-t')$, then the noise amplitude
$N_0$ will satisfy

\begin{equation}
N_0 
= \tilde{{\cal N}} (0)
=  2 T \tilde{\Gamma}(0) \;,
\label{Tdissip} 
\end{equation}
where we have used Eq. (\ref{fluctdiss2}) and taken the limit as $\omega\to
0$.  Note in particular from Eq. (\ref{Tdissip}) that there is no $T=0$ local
limit for the noise correlation function and, thus, no local limit for the
dissipation terms in the effective EOM (\ref{full eqmotion2}) at $T=0$.

Equation (\ref{Tdissip}) is one of our main results. This simple relationship
between the noise and dissipation has appeared previously as a high
temperature limit in various applications, but our derivation is quite general
and depends only on the localizability of the noise and dissipation. We shall
describe in the next subsection how, and under what circumstances, the
localization in time is realized.

The noise amplitude $\tilde{{\cal N}} (0)$ can be determined as follows.  The
inverse transform of the one-loop and two-loop noise kernels Eqs.
(\ref{noise1kernel}) and (\ref{noise2kernel}) is defined by

\begin{equation}
{\cal F}^{-1} [\tilde{{\cal N}}_i ({\bf p}, \omega)] = 
{\cal N}_i({\bf x}-{\bf x}',t-t') 
= \int \frac{d^3 p}{(2 \pi)^3} \int_{-\infty}^{+\infty} 
\frac{d \omega}{(2 \pi)}  \exp\left[-i {\bf p} \cdot ({\bf x}-{\bf x}') 
+ i \omega (t-t')
\right] \tilde{{\cal N}}_i ({\bf p}, \omega)\;.
\label{Nitrans}
\end{equation}

\noindent
{}For homogeneous fields, $\varphi_c(x) \equiv \varphi_c(t)$, all loop
diagrams in the effective action carries zero external space momentum, ${\bf
  p} =0$, and then immediately follows from Eq.  (\ref{Nitrans}) that ${\cal
  N}_1({\bf x}-{\bf x}',t-t')$ and ${\cal N}_2({\bf x}-{\bf x}',t-t')$ are
delta correlated in space ($ {\cal N}_i({\bf x}-{\bf x}',t-t') \sim
\delta({\bf x} - {\bf x}')$).  If also a local limit for the noise two-point
correlations are valid, then the noise amplitudes become, from Eqs.
(\ref{noise1kernel}) and (\ref{noise2kernel}), respectively

\begin{equation}
\tilde{{\cal N}}_1 (0) = 2 g^4 \int \frac{d^3 k}{(2 \pi)^3} 
\int_{-\infty}^{+\infty} \frac{d \omega_1}{2 \pi}
n(\omega_1) \left[1+ n(\omega_1)\right] 
\tilde{\rho}_\chi^2({\bf k},\omega_1)\;,
\label{N1tilde}
\end{equation}
while for the two-loop noise kernel,

\begin{eqnarray}
\tilde{{\cal N}}_2 (0) &=& 
2 g^4 \int \frac{d^3 k}{(2 \pi)^3} \frac{d^3 q}{(2 \pi)^3}
\int_{-\infty}^\infty {d\omega_1 \over 2\pi}{d\omega_2 \over 2\pi}
n(\omega_1)n(\omega_2) [1+n(\omega_1+\omega_2)] \nonumber \\
&\times& 
\tilde{\rho}_\chi({\bf k},\omega_1) \tilde{\rho}_\chi({\bf q},\omega_2)
\tilde{\rho}_\phi({\bf k}+{\bf q},\omega_1+\omega_2)\;.
\label{N2tilde}
\end{eqnarray}

\noindent
{}From now on we will assume the simplest case of homogeneous fields and study
the localization for the noise kernels in time.  If such a local approximation
exists for the noise kernels, then Eq. (\ref{Tdissip}) suggests that the
{}Fourier ($\omega$) transform, $\tilde{{\cal N}}_j (\omega)$, should not
differ much from the $\omega=0$ value, at least for some region in $\omega$
space, $-\omega_{\rm local}\lesssim \omega \lesssim \omega_{\rm local}$. We
now try to find this region in $\omega$ space.

\subsection{The one-loop noise kernel}

{}For the general case, which should be valid for both the low as well high
temperature limits, we can use the full expression for the spectral density,
Eq. (\ref{rho}), which in terms of the decay width $\tau_i^{-1} = - {\rm Im}
\Sigma_i/[4 \omega_i({\bf k})]$, can be written in the form ($i=\phi,\chi$)

\begin{equation}
\tilde{\rho}_i({\bf k}, \omega) = \frac{4 \omega_i({\bf k}) 
\tau_i^{-1}({\bf k},\omega)}{ \left[ \omega^2 - \omega_i^2({\bf k})
\right]^2 + \left[ 2  \omega_i({\bf k}) 
\tau_i^{-1}({\bf k},\omega) \right]^2 }\;.
\label{rhotau}
\end{equation}

\noindent
We will now study the localization of the one-loop noise kernel Eq.
(\ref{N1}). Recall that here we are restricting our analysis to the parameter
regime where $\lambda \ll g^2$, where the first term in Eq. (\ref{N1}) can be
neglected compared to the second one. The second term will only involve the
spectral density for the $\chi$ field.  The decay width $\tau_\chi^{-1}$
receives contributions from both one and two $\sigma$ loop diagrams. Both
results are well known (see e.g. Ref. \cite{mx} for the first one and Ref.
\cite{bgr} for the second). The single $\sigma$ loop diagram is ${\cal
  O}(h^2)$, while the two-loop diagram is ${\cal O} (g^4,h^2)$. Therefore, for
parameter values such that $h \gg g^2$ (these parameter values were in fact
the ones shown to be relevant for the dynamics studied in Refs.
\cite{bgr,bgr2,br,br2}), the dominant contribution comes from the single loop.
Thus, for the dominant decay channel we have that ($k\equiv |{\bf k}|$)
\cite{mx}

\begin{eqnarray}
\tau_\chi^{-1}({\bf k},\omega) 
\simeq
\frac{h^2 {\cal M}^2}{16 \pi \omega_\chi({\bf k})}
\left[ \theta(\omega-k)-\theta(-\omega-k) \right]
+\frac{h^2 {\cal M}^2}{8 \pi \omega_\chi({\bf k})}
\frac{T}{k}
\ln\left[
{1-e^{-\beta|\omega+k|/2}\over 1-e^{-\beta|\omega-k|/2}} \right]
\label{tauchi}\:, 
\end{eqnarray}
which is valid for very light $\sigma$ bath fields, $m_\sigma \approx 0$.

Equation (\ref{noise1kernel}) then becomes (for $p\equiv |{\bf p}|=0$)

\begin{eqnarray}
\tilde{{\cal N}}_1 (\omega) &\simeq& 
g^4\left(e^{\beta\omega}+1\right)
\int \frac{d^3 k}{(2 \pi)^3} \int_{-\infty}^\infty {d\omega' \over 2\pi}
n(\omega') n(\omega-\omega')
\tilde{\rho}_\chi({\bf k},\omega') 
\tilde{\rho}_\chi({\bf k},\omega-\omega')\;.
\label{N1G_1}
\end{eqnarray}

\subsubsection{The low temperature limit}

At low temperatures $T \ll m_\chi$, we can easily verify that the dominant
contributions to the noise kernel Eq.  (\ref{noise1kernel}) come from
$k,\omega \ll m_\chi$.  We can approximate the spectral function Eq.
(\ref{rho}) for the $\chi$ field as

\begin{equation}
\rho_\chi \simeq \frac{4}{m_\chi^3 }\tau_\chi^{-1}\:,
\label{rholowT}
\end{equation}
with $\omega_\chi({\bf k}) \sim m_\chi$ in Eq. (\ref{tauchi}).  Thus, from
Eqs. (\ref{N1}), (\ref{noise1kernel}) and (\ref{rholowT}), and for the same
parameter values considered before, $h \gg f,g^2$ and $\lambda \ll g^2$, we
obtain

\begin{eqnarray}
\tilde{{\cal N}}_1 (\omega) &\simeq& 
\frac{16 g^4}{ m_\chi^6} \left(e^{\beta\omega}+1\right)
\int \frac{d^3 k}{(2 \pi)^3} \int_{-\infty}^\infty {d\omega' \over 2\pi}
n(\omega') n(\omega-\omega')
\tau_\chi^{-1}({\bf k},\omega') 
\tau_\chi^{-1}({\bf k},\omega-\omega')\;.
\label{N1G_2}
\end{eqnarray}
The dimensionful parameters can all be rescaled out of the integral to leave

\begin{equation}
\tilde{{\cal N}}_1 (\omega)=g^4h^4
{{\cal M}^4T^4\over m_\chi^8} {\rm F}(h,\beta\omega)\;.
\label{Fdef}
\end{equation}

\noindent
In principle ${\rm F}$ only depends on $\beta\omega$, but in practice there is
a breakdown of (\ref{rholowT}) at $\omega=k$ when the light field mass
$m_\sigma= 0$, resulting in the $h$ dependence.

\begin{figure}[ht]
  \vspace{1cm} \epsfysize=7cm {\centerline{\epsfbox{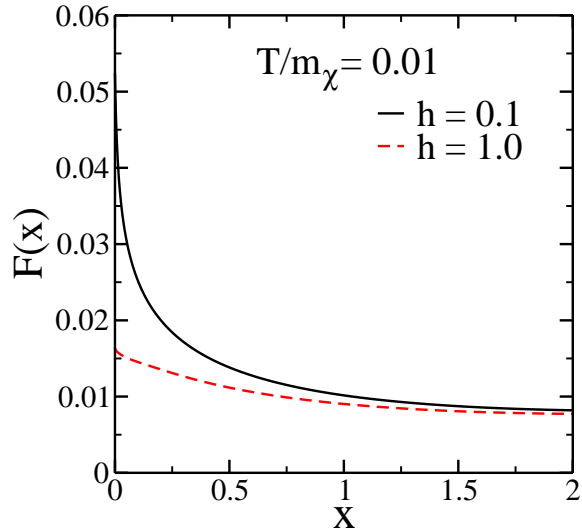}}}
\caption{The rescaled noise amplitude ${\rm F}(h,x)$, where $x=\omega/T$, 
  at $T/m_{\chi}=0.01$ for $h=0.1$ (solid) and $1.0$ (dashed). Below this
  temperature these curves do not change significantly.  }
\label{fig1}
\end{figure}

The results of a numerical evaluation of the function ${\rm F}(h,\beta
\omega)$ have been plotted in {}Fig. \ref{fig1}, which takes the temperature
$T/m_{\chi}=0.01$ for the cases $h=0.1$ (solid) and $1.0$ (dashed).  Below
this temperature, the numerical calculations find almost no change in ${\rm
  F}(h,\beta \omega)$, which is consistent with our parameterization.  The
best 3-parameter fit to the numerical curve at small values of $x=\omega/T$ is
of the form

\begin{equation}
{}{\rm F}(h,x)=A(h)+B(h) \,x\ln(x)+C(h) \,x \;,
\end{equation}

\noindent
where we find $A(0.1)=0.0458$, $B(0.1)=0.144$ and $C(0.1)=0.126$ and
$A(1.0)=0.0161$, $B(1.0)=0.0068$ and $C(1.0)=-0.0004$.  These numerical
results provide convincing evidence that the noise kernel can be localized in
the sense described earlier. {}For example, we can say for $h=0.1$ that the
noise kernel is localized on a timescale $1600T^{-1}$ with 95\% accuracy and
for $h=1.0$ the noise kernel is localized on a timescale $100T^{-1}$ with 95\%
accuracy. Note that, although this may be a large timescale compared to the
scale of particle interactions, it may still be a small timescale compared to
the timescale of a cosmological application.

It is interesting that the numerical evidence supports the conclusion that the
noise kernel is not analytic in $\omega$. The logarithmic terms in $\omega$
are related to subleading non-local effects. This also has an important
implication for the dissipation terms. The dissipation kernels ${\cal C}_\chi$
and ${\cal D}_\chi$ are related to the noise kernel by the
fluctuation-dissipation relation Eq. (\ref{fluctdiss2}), and in the low
temperature limit,

\begin{equation}
\tilde {\cal C}_\chi=-i\omega\tilde{\cal D}_\chi=
-2i\beta\omega\tilde{\cal N}.
\end{equation}
Therefore the dissipation terms can have a local analytic expansion at leading
order to give a $\dot\varphi$ term, but beyond that there is not a local
analytic expansion but rather terms of the form

\begin{equation}
\ln\left({\partial\over\partial t}\right)
{\partial^2\over\partial t^2}\varphi \;.
\end{equation}
{}For a definition of the logarithmic derivative, see Ref. \cite{Moss:2004dp}.

\subsubsection{The high temperature limit}

\begin{figure}[ht]
  \vspace{1cm} \epsfysize=7cm {\centerline{\epsfbox{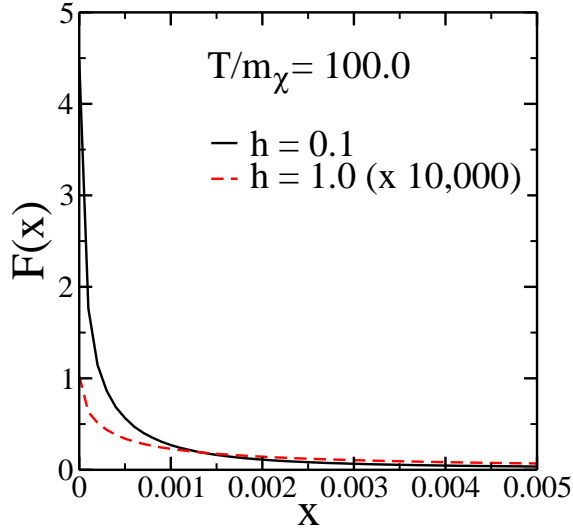}}}
\caption{The rescaled noise amplitude ${\rm F}(h,x)$, where $x=\omega/T$,
  at $T/m_{\chi}=100.0$ for $h=0.1$ (solid) and $1.0$ (dashed, increase by a
  factor $10,000$).}
\label{fig2}
\end{figure}

{}For high temperature $T\gg m_\chi$ and $T\gg\omega$, the specral function
$\tilde\rho_\chi$ is concentrated in the range of values of $\omega'$ close to
$\pm\omega({\bf p})$, where

\begin{equation}
\omega_\chi({\bf k})=(k^2+m_\chi^2)^{1/2}.
\end{equation}
We take the distribution functions out of the energy integral,

\begin{equation}
\tilde{{\cal N}}_1(\omega)\approx -g^4(e^{\beta\omega}+1)
\int {d^3k\over (2\pi)^3}n(\omega_\chi({\bf k}))
\left[1+n(\omega_\chi({\bf k}))\right]
\int_{-\infty}^\infty{d\omega'\over 2\pi}
\tilde\rho_\chi({\bf k},\omega')
\tilde\rho_\chi({\bf k},\omega-\omega')\;.
\end{equation}
The spectral function is given by Eq. (\ref{rhotau}), with
$\tau_\chi(k)\equiv\tau_\chi({\bf k},\omega_\chi({\bf k}))$.  After the
$\omega'$ integration,

\begin{equation}
\tilde{{\cal N}}_1(\omega)\approx g^4
\int {d^3k\over (2\pi)^3}
{4\tau_\chi(k)\over 4+\tau_\chi(k)^2\omega^2}
{n(\omega_\chi({\bf k}))\left[ 1+n(\omega_\chi({\bf k})) \right]
\over\omega_\chi({\bf k})^2}.
\end{equation}
Note that this result for the noise kernel is consistent with the earlier
relationship ${\tilde{\cal N}}_1(0)=2T{\tilde\Gamma}(0)$, where
$\tilde{\Gamma}(0)$ is the high temperature dissipation coefficient first
calculated by Hosoya and Sakagami \cite{hs}.

The noise kernel is localized in the sense which we defined earlier. For a
rough estimate, we introduce the relaxation timescale $\tau_\chi$ at $k = T$
and take $\tau_\chi$ out of the $k$ integral. The noise kernel is then
localized at the timescale $\sim 30\tau_\chi$ with accuracy 99\%.
Alternatively, we can say in less formal terms that the noise kernel appears
localized for physics on timescales larger than the relaxation time.  A
numerical calculation of the noise amplitude is shown in Fig. \ref{fig2} and
it is consistent with these analytic estimates of the localization scale.

\subsection{The two-loop noise kernel}

{}For the two-loop noise kernel, the expression analogous to Eq.
(\ref{N1G_1}), from Eq. (\ref{noise2kernel}), is

\begin{eqnarray}
\tilde{{\cal N}}_2(\omega)
&=& g^4\left(e^{\beta\omega}+1\right)
\int \frac{d^3 k}{(2 \pi)^3} \frac{d^3 q}{(2 \pi)^3}
\int_{-\infty}^\infty {d\omega_1 \over 2\pi}{d\omega_2 \over 2\pi}
n(\omega_1)n(\omega_2) n(\omega-\omega_1-\omega_2) \nonumber \\
&\times& 
\tilde{\rho}_\chi({\bf k},\omega_1) \tilde{\rho}_\chi({\bf q},\omega_2)
\tilde{\rho}_\phi({\bf k}+{\bf q},\omega-\omega_1-\omega_2)\;,
\label{N2G_1}
\end{eqnarray}

\noindent
where $\tilde{\rho}_\chi$ is given by Eq. (\ref{rhotau}) with decay width
given by Eq. (\ref{tauchi}), and $\tilde{\rho}_\phi$ is given by an equivalent
expression with $\tau_\phi$ in place of $\tau_\chi$. {}For the range of
parameters discussed above, the dominant decay channel for $\phi$ in the
thermal bath is the two-loop channel $\phi \to \phi \chi \chi$. The explicit
expression for this decay term (and for any of the two-loop terms in general)
is a very complicated one and there is no simple result available for it,
unfortunately.  An explicit expression in terms of momenta integrals was
derived in \cite{bgr}, for the on-shell case $\omega = \omega({\bf p})$ only.
However, we note that for the dominant decay channel $\phi \to \phi \chi
\chi$, $\tau^{-1}_\phi \sim {\cal O}(g^4)$, and since in Eq. (\ref{N2G_1}) we
already have a $g^4$ factor and in each of the $\tilde{\rho}_\chi$ terms we
have an extra factor of $h^2$ (from Eq. (\ref{tauchi})), then $\tilde{{\cal
    N}}_2 \sim {\cal O} (g^8 h^4)$ and it is subleading compared to the
one-loop kernel $\tilde{{\cal N}}_1$. Because of this, there is no need to
proceed further with an analysis of the two-loop term here.

\section{Conclusions}

The effective (microscopic) equations of motion of systems coupled to an
environment in the form of a heat bath typically include nonlocal dissipation
and noise (stochastic) kernels. These terms are generally complicated and need
to be dealt with somehow for practical purposes for studying the dynamical
evolution of these systems.  The same is true for the evolution of quantum
fields in a medium.  In this paper we have studied and approached the problem
of localization of the nonlocal kernels in the context of the dynamics of a
scalar background field coupled to an environment made of other scalar fields.
We have paid special attention to a particular model where dissipation emerges
as a consequence of a two-stage decay mechanism where a scalar field decays
into a light one through a catalyst heavy field.  This model was identified in
Refs. \cite{br,br2} and can emerge naturally in the context of Grand Unified
and Supersymmetric theories, with applications in several contexts, like
inflationary cosmology and particle physics dynamics in general.

{}For this model, we have studied in this paper the nonlocal dissipation and
noise kernels that emerge in general in the microscopic derivation of the
effective equations of motion for quantum fields. From this we have derived a
fluctuation-dissipation theorem and determined the appropriate conditions for
when the kernel terms can be well approximated by a local form.

We should point out that small differences between the local and nonlocal 
kernels over large evolution timescales can still manifest themselves as a
secular drift. This, if the scalar field is oscillating, can lead
to a phase shift that might accumulate, but does not affect the important
physical quantities like the rate of dissipation of energy. If the dissipation
terms dominate over the other kinetic terms, then the overdamped
solution emerges and a small correction to the equations is equivalent to a 
shift in the origin of time, which has no physical significance.

{}For a scalar field with a expectation value $\varphi_c$ we have found that,
in certain situations when the local form is possible, the effective equation
of motion reduces to

\begin{equation}
\left[ \Box + M_\phi^2 \right]\varphi_c (x) +
2\Gamma(\varphi,T)\varphi_c^2(x)\dot\varphi_c(x)
+\frac{\lambda}{3 !} \varphi_c^3 (x)=\varphi_c(x)\xi(x) ,
\end{equation}
where the stochastic noise source has a correlation function

\begin{equation}
\langle \xi(x)\xi(x')\rangle=2\Gamma(\varphi,T) T\,\delta(x-x') .
\end{equation}
The situations which we have examined have been for approximately homogeneous
fields coupled to the heat bath through a catalyst field with coupling $g$ and
weak self-coupling $\lambda=O(g^2)$:

\begin{enumerate}
  
\item When the catalyst mass is large compared to the temperature and the
  scalar field is slowly varying on a timescale $CT^{-1}$, where $C$ is a
  constant which was determined numerically.
  
\item When the catalyst mass is small compared to the temperature and the
  scalar field is slowly varying on the relaxation timescale of the catalyst
  field.

\end{enumerate}
In case 1, we have also given numerical evidence that the dissipative term is
localized on a specified timescale but beyond leading order it possesses no
analytic expansion with respect to the time coordinate, since beyond this
order it is dominated by logarithmic terms.

{}Finally, we would like to point out a similar model to the one
we analyzed here that was studied by the authors in Ref. \cite{felder}.
In that case, they have studied the process of how the homogeneous field
$\phi$ can produce fluctuations of $\chi$ through the process of
parametric resonance. In that case, oscillations of the scalar field $\phi$ 
lead to the growth of the effective mass of $\chi$, which can then cause it 
to eventually decay into excitations of $\sigma$, even if $m_\sigma$ is 
much greater than the vacuum mass of $\phi$ or $\chi$. Though they analyze
a different dynamical regime for the scalar field $\phi$, we think that this
process of energy transfer from the field $\phi$ to $\sigma$, through
parametric resonance, is a nice counterpart to the quantum process we have
studied here.

\acknowledgments

This work was partially supported by a U.K. Particle Physics and Astronomy
Research Council (PPARC) Visiting Researcher Grant.  In addition A.B. is
partially supported by PPARC and R.O.R. is partially supported by Conselho
Nacional de Desenvolvimento Cient\'{\i}fico e Tecnol\'ogico (CNPq - Brasil)
and Funda\c{c}\~ao de Amparo \`a Pesquisa do Estado do Rio de Janeiro
(FAPERJ).


\begin{thebibliography}{99}
%
  
\bibitem{im} I. G. Moss, Phys. Lett. B{\bf 154}, 120 (1985).
  
\bibitem{wi} A. Berera, Phys. Rev. Lett. {\bf 75}, 3218 (1995); Phys. Rev. D
  {\bf 54}, 2519 (1996).
  
\bibitem{tb} A. N. Taylor, and A. Berera, Phys. Rev. D {\bf 62}, 083517
  (2000).
  
\bibitem{hmb} L. M. H. Hall, I. G. Moss, and A. Berera, Phys. Rev. D {\bf 69},
  083525 (2004); Phys. Lett. B {\bf 589}, 1 (2004).
  
\bibitem{xg} Z. Xu and C. Greiner, Phys. Rev. D{\bf 62}, 036012 (2000).
  
\bibitem{bv} D. Boyanovsky and H. J. de Vega, hep-ph/9909372.
  
\bibitem{hs} A. Hosoya and M. Sakagami, Phys. Rev. D{\bf 29}, 2228 (1984).
  
\bibitem{morikawa} M. Morikawa and M. Sasaki, Phys. Lett. B{\bf 165}, 59
  (1985); M. Morikawa, Phys. Rev. D{\bf 33}, 3607 (1986).
  
\bibitem{ring1} A. Ringwald, Ann. Phys. (N.Y.) {\bf 177}, 129 (1987).
  
\bibitem{boya} D. Boyanovsky and H. J. de Vega, Phys. Rev.  D{\bf 47}, 2343
  (1993).
  
\bibitem{GR} M. Gleiser and R. O. Ramos, Phys. Rev. D{\bf 50}, 2441 (1994).
  
\bibitem{cooper} F. Cooper, S. Habib, Y. Kluger, E. Mottola, J. P. Paz and P.
  R. Anderson, Phys. Rev. D{\bf 50}, 2848 (1994).
  
\bibitem{bgr} A. Berera, M. Gleiser and R. O. Ramos, Phys. Rev. {\bf D58},
  123508 (1998).

  
\bibitem{bgr2} A. Berera, M. Gleiser and R. O. Ramos, Phys. Rev. Lett.  {\bf
    83}, 264 (1999).
  
\bibitem{br} A. Berera and R. O. Ramos, Phys. Rev. D {\bf 63}, 103509 (2001).

  
\bibitem{br2} A. Berera and R. O. Ramos, Phys. Rev. D {\bf 71}, 023513 (2005).
  
\bibitem{mx} I. G. Moss and C. Xiong, hep-ph/0603266.
  
\bibitem{bb4} M. Bastero-Gil and A. Berera, Phys. Rev. D {\bf 76},
043515 (2007).
  
\bibitem{schw}J. Schwinger, Jour. Math. Phys. (N.Y.) {\bf 2}, 407 (1961); P.
  M. Bakshi and K. T. Mahanthappa, Jour. Math. Phys. (N.Y.) {\bf 4}, 1 (1963);
  {\bf 4}, 12 (1963); L. V. Keldysh, Sov. Phys. JETP {\bf 20}, 1018 (1964); 
A.  Niemi and G. Semenoff, Ann. Phys. (NY) {\bf 152}, 105 (1984);
  Nucl. Phys. {\bf B230}, 181 (1984).
  
\bibitem{KKR}T. Koide, G. Krein and R. O. Ramos, Phys. Lett.  {\bf B636}, 96
  (2006).
  
\bibitem{dh} A. Das and M. Hott, Phys. Rev. D{\bf 50}, 6655 (1994).
  
\bibitem{im2} I. G. Moss, Nucl. Phys. B{\bf 631}, 500 (2002).
  
\bibitem{aarts} G. Aarts and A. Tranberg, Phys. Lett. B{\bf 650}, 65 (2007). 
  
\bibitem{chou}K. Chou, Z. Su, B. Hao and L. Yu, Phys. Rep. {\bf 118}, 1
  (1985).
  
\bibitem{rivers}An accessible introduction to the real-time formalism can be
  found in R. Rivers, {\it Path Integral Methods in Quantum Field Theory},
  Cambridge University Press, (Cambridge 1987).
  
\bibitem{weert}N. P. Landsman and Ch. G. van Weert, Phys. Rep.  {\bf 145}, 141
  (1987).
  
\bibitem{yoko} J. Yokoyama, Phys. Rev. D{\bf 70}, 103511 (2004).
  
\bibitem{gert}G. Aarts and J. Berges, Phys. Rev. D {\bf 64}, 105010 (2001).
  
  
\bibitem{lebellac}M. Le Bellac, {\it Thermal Field Theory} (Cambridge
  University Press, Cambridge, 1996).
  
\bibitem{Moss:2004dp} {I.~G.} {Moss} {and} J.~P. Norman, Phys. Rev. D {\bf
    70}, 103506 (2004).

\bibitem{felder}G. Felder, L. Kofman and A. Linde, Phys. Rev. D {\bf 59},
123523 (1999).


\end{thebibliography}
\end{document}